\documentclass[12pt, preprint]{aastex}

\newcommand{\age}{\gtrsim}
\newcommand{\ale}{\lesssim}
\newcommand{\ba}{\begin{eqnarray}}
\newcommand{\ea}{\end{eqnarray}}
\newcommand{\bas}{\begin{eqnarray*}}
\newcommand{\eas}{\end{eqnarray*}}
\newcommand{\be}{\begin{equation}}
\newcommand{\ee}{\end{equation}}
\newcommand{\bes}{\begin{equation*}}
\newcommand{\ees}{\end{equation*}}
\newcommand{\bft}{\begin{figure}[t]}
\newcommand{\bfh}{\begin{figure}[h]}
\newcommand{\bfb}{\begin{figure}[b]}
\newcommand{\bd}{\begin{displaymath}}
\newcommand{\ed}{\end{displaymath}}

\begin{document}
\title{Alfv\'en Waves and Turbulence in the Solar Atmosphere and Solar
Wind}
\author{Andrea Verdini \altaffilmark{1} \and Marco Velli \altaffilmark{2}}
\altaffiltext{1}{Dipartimento di Astronomia
        e Scienza dello Spazio, Universit\`a degli Studi di Firenze, 
        Firenze, 
        Italy}
\altaffiltext{2}{JPL, California Institute of Technology, Pasadena, California,
        US.; on leave of absence from the Dipartimento di Astronomia e Scienza
dello Spazio, Universit\`a degli Studi di Firenze, Italy}
\begin{abstract}
We solve the problem of propagation and dissipation of Alfv\'enic turbulence in a model solar atmosphere consisting of a static photosphere and chromosphere, transition region, and open corona
and solar wind, using a phenomenological model for the turbulent dissipation based on wave reflection.
We show that most of the dissipation for a given wave-frequency spectrum occurs in the lower corona, and the overall rms amplitude of the fluctuations evolves in a way consistent with observations. The frequency spectrum, for a Kolmogorov-like slope, is not found to change dramatically from the photosphere to the solar wind,  however it does preserve signatures of transmission throughout the lower atmospheric layers, namely oscillations in the spectrum at high frequencies reminiscent of the resonances found in the linear case.  These may disappear once more realistic couplings for the non-linear terms are introduced, or if time-dependent variability of the lower atmospheric layer is introduced.
\end{abstract}
\keywords{MHD -- waves -- turbulence -- Sun: solar wind}

\section{INTRODUCTION}
In situ measurement of magnetic and velocity field fluctuations from Helios and Ulysses 
have revealed a
broad developed spectrum for frequencies ranging from $10^{-4}~\mathrm{Hz}$
to $10^{-2}~\mathrm{Hz}$.
Typically, 
a strong correlation between magnetic field and velocity fluctuations in this distance range
persists \citep{Mangeney_al_1991} corresponding to an outwardly propagating
spectrum.
It is well known that nonlinear terms couple Alfv\'en waves
propagating 
in opposite directions. Also, the basic nonlinearity in homogeneous MHD in the presence of a majority of one type of waves forces the evolution with time to increase the dominance, preferentially dissipating the minority component in a process called dynamical alignment (Veltri et al. 1980) which is not observed in the solar wind. Therefore the presence of a well-developed spectrum 
together with a preferred direction of propagation has remained a mystery.
The question that naturally arises therefore concerns the drivers for the continuing and anomalous (compared to homogeneous MHD predictions)
nonlinear cascade in this outwardly dominant case.
Among the possible drivers of a nonlinear cascade in the solar atmosphere 
are compressible effects, which couple Alfv\'en
waves with slow and fast modes, or couplings due to the strong 
gradients in the atmosphere. Among the first are phenomena
such as parametric decay 
\citep{Pruneti_Velli_1997, DelZanna_al_2001} and wave-steepening
\citep{Suzuki_Inutsuka_2005a}. 
Gradients transverse to the mean
magnetic field directions lead to phase-mixing, i.e. development of 
small scales in directions perpendicular to that of propagation. Finally,
the gradients due to stratification cause wave-reflection,
which naturally produces the waves propagating in opposite direction 
required for the classical incompressible cascade, as first suggested 
by \citet{Velli_al_1989}. Disentangling the role of all of these processes at 
once would require fully 3D calculations in a realistic atmosphere model, 
a feat beyond present numerical capabilities. We therefore focus here 
on the role of wave-reflection, which has been extensively studied in 
the linear case 
(\citealp{Heinemann_Olbert_1980, Leroy_1980,
Hollweg_1978a} among the first) while less so in the
nonlinear one \citep{Matthaeus_al_1983, Matthaeus_al_1994}.\\ 

Some constraints on the frequency spectrum and the energies for the outward and
inward propagating components are derived from the observations.
The Alfv\'enic fluctuation power spectrum in the fast solar wind
evolves with distance ($R$) not self-similarly with a power-law
dependence on $\omega$ with slope -1 and -5/3 at low and high
frequencies respectively. 
The two intervals are separated by a critical frequency
($\omega_*$) which depends on $R$.
Identifying the fluctuations with Alfv\'en waves it is useful to adopt  the Els\"{a}sser variables 
${\bf z}^\pm={\bf u}\mp\textrm{sign}(B_0){\bf b}/\sqrt{4\pi\rho}$ 
(corresponding respectively to
outward and inward propagating Alfv\'en waves if the mean magnetic field 
$B_0$ is pointing outward from the sun)
The energy per unit mass residing in the outward and inward propagating
modes ($E^\pm=|{\bf z}^\pm|^2$ respectively) both decrease with distance and
for $R<2.5~\mathrm{AU}$ $E^+\propto R^{-1.48}$ and
$R^-\propto E^{-0.42}$ \citep{Bavassano_al_2000a}. 
The normalized cross helicity,
$\sigma_c=(E^+-E^-)/(E^++E^-)$ which accounts for the imbalance between the 
outward and inward component, also evolves with distance and it is 
approximately equal to one in the inner solar wind, it decreases 
for $R>0.4~\mathrm{AU}$ and oscillates around $\approx 0.4$  for 
$R>2.5~\mathrm{AU}$ \citep{Bavassano_al_2000b}.\\

It must be recalled that 
the observed features can not be explained either by linear propagation theory
(including reflection) or by MHD turbulence separately.\\
A linear analysis applied to the solar wind case 
shows that low frequency waves
($\omega<10^{-5}~\mathrm{Hz}$) experience the strongest reflection in the
photosphere, chromosphere and corona \citep{Hollweg_1978a, Hollweg_1981,
Similon_Zargham_1992}. 
Their flux
at the transition region is greatly reduced
(even if a considerable power is transmitted to the corona) 
and in the outer (supersonic) solar wind the radial dependence of
$\sigma_c$ is similar to the observed at higher frequencies
\citep{Velli_al_1991}.\\ 
On the other hand the high frequency waves
($10^{-4}~\mathrm{Hz}<\omega<10^{-2}~\mathrm{Hz}$) are almost completely
transmitted (even if in the photosphere and chromosphere their reflection is
relatively high), both the $E^+$ and $E^-$ energies decrease faster than the scaling
observed and finally $\sigma_c\approx1$ in the outer solar wind
\citep{Velli_al_1991}.\\
The dynamics of a well developed turbulent state in the expanding solar wind 
has been studied as well 
and ordering of the characteristic time scale 
which should effectively favor the development of a turbulent cascade in planes
perpendicular to the direction of wave propagation 
(along the magnetic field) has been found 
\citep{Zank_al_1996, Matthaeus_al_1998, Matthaeus_al_1999, 
Dmitruk_al_2001b, Dmitruk_al_2002, Dmitruk_Matthaeus_2003,
Oughton_al_2001, Oughton_al_2004}.
Numerical models capable of reproducing the observed $\sigma_c$ profiles in
the supersonic part of the solar wind 
\citep{Zhou_Matthaeus_1989,Zhou_Matthaeus_1990} 
or the spectral evolution
\citep{Tu_al_1984, Tu_1988, Velli_al_1989} necessarily use
ad hoc assumption and simplification, and, even if considerable 
advances have been made,
a complete understanding of
the solar turbulent spectrum and the solar wind acceleration 
\citep{Li_al_1999, Habbal_al_1995}
has not been achieved.\\

Here, we investigate the combined effect of wave reflection and turbulent 
dissipation in order to understand the relative importance of linear and 
nonlinear effects on the overall evolution of the fluctuation amplitudes.
Comparison of the numerical results with some observations give some
constrains on the fields at the photospheric and coronal level for which
data are still missing with implications for numerical models of solar wind
acceleration.\\
In the context of a reflection driven turbulent cascade process
another interesting issue concernes the evolution of the turbulent spectrum.
If one supposes
that the Alfv\'en waves are injected at the photospheric base at a well
defined frequency or with a given correlation time one would expect to find a
signature of this characteristic time-scale in the observed spectrum at
$1~\mathrm{AU}$ 
(or in other words, one can ask if discrete modes and turbulence can coexist
\citep{Dmitruk_al_2004}).
No injection frequency is observed in the solar wind spectrum so one can ask
if both the
turbulent evolution and the frequency-dependent transmission properties of
the solar atmosphere and wind can efficiently smooth this supposed strong
forcing signature.\\

We integrate
numerically the equations for the velocity and magnetic field fluctuations
(written in terms of the Els\"{a}sser fields)
for a stationary model atmosphere with a
supraspherically expanding wind, 
form the photosphere to one AU for a set of frequency
chosen in the range 
$10^{-6}~\mathrm{Hz}<\omega<10^{-2}~\mathrm{Hz}$.
Each wave is identified via its frequency while a phenomenological nonlinear
term is added to the equations in order to account for both turbulent
dissipation and frequency coupling.\\

\section{THE MODEL}\label{sec.model}
The equations describing the propagation of Alfv\'en waves in an
inhomogeneous stationary medium can be derived from the Magnetohydrodynamic
equations (MHD) under the hypotheses of incompressible adiabatic transverse
fluctuations. The velocity $({\bf u})$
and magnetic field fluctuations $({\bf b})$ can be combined to form the
Els\"{a}sser variables 
${\bf z}^\pm={\bf u}\mp{sign({\bf B_0}){\bf b}}/{\sqrt{4\pi\rho}}$
which describe Alfv\'en waves propagating outward $({\bf z^+})$ or
inward $({\bf z^-})$. ${\bf B_0}$ stands for the average magnetic
field 
while $\rho$ is the mass density. 
In terms of these variables
the equations for the two fields read:
\bd
\frac{\partial {\bf z}^\pm}{\partial t}+[({\bf U}\pm {\bf V_a})\cdot\nabla]
{\bf z}^\pm+({\bf z}^\mp\cdot\nabla)({\bf U}\mp {\bf V_a})+
\ed
\be 
\pm\frac{1}{2}({\bf z^\mp -z^\pm})[\nabla\cdot
{\bf V_a}\mp\frac{1}{2}(\nabla\cdot {\bf U})]=
-({\bf z}^\mp\cdot\nabla){\bf z}^\pm,
\ee\label{eq.Alf_wind} 
where ${\bf U}$ is the mean wind speed and the Alfv\'en speed is ${\bf
  V_a}={\bf B_0}/\sqrt{4\pi\rho}$, co-linearity between magnetic and
gravitational field is assumed. On the right hand side we have grouped the
nonlinear terms (except the total pressure, which in
the limit of incompressible fluctuations can also be written as
the product of ${\bf z}^+$, ${\bf z}^-$ and their gradients).
In the linear part of the eq.~\ref{eq.Alf_wind} we can recognize a propagation term
(II) and two terms accounting for reflection due to the variation of
the properties of the medium, one isotropic (IV) while the other (III)
involves variations along the fluctuations' polarization.\\

\clearpage
\begin{figure*}[t]
\plotone{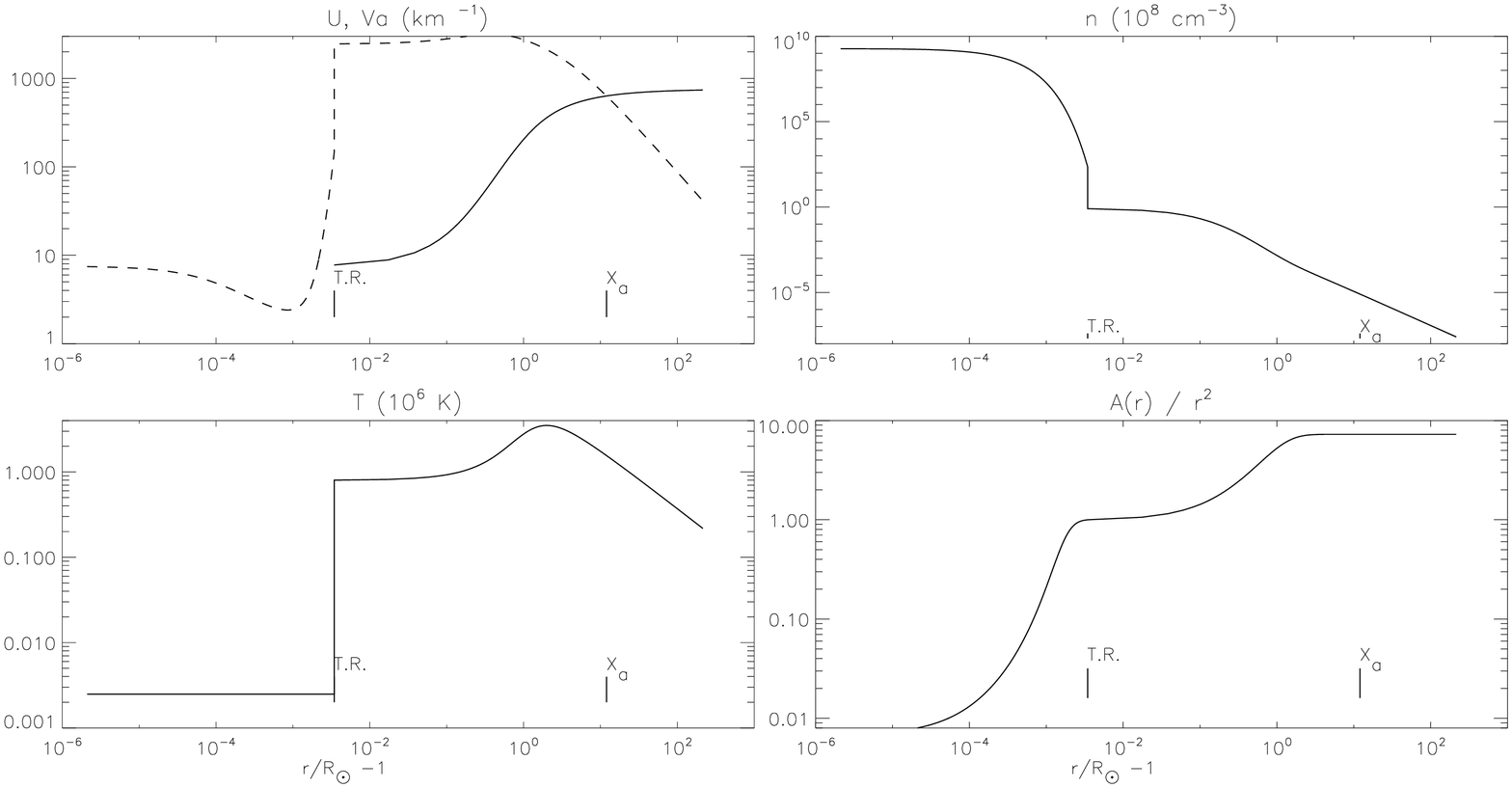}
\caption{From top left, clockwise:
wind speed (solid line) and Alfv\'en speed (dotted line),
numerical density, temperature and expansion factor
as function of heliocentric distance for the modeled atmosphere.}
\label{fig.atmosp}
\end{figure*}
\clearpage
The chromosphere and the photosphere are modeled as a static layer,
$2400~\mathrm{km}$ thick, with the magnetic field organized in flux tube in
supra-spherical geometry with constant temperature. The density varies almost
exponentially and the magnetic field varies according to the flux tube
expansion ($A$) in order to reproduce the properties of a coronal
hole in the quiet Sun \citep{Hollweg_al_1982a}. 
Across the transition region the density falls off by two orders of
magnitude, the wind passes from a speed of $0~\mathrm{km~s^{-1}}$ to
$8~\mathrm{km~s^{-1}}$ while the
magnetic field strength is continuous (about $10~\mathrm{G}$).
The corona also expands supra-spherically and its temperature profile is chosen
to fit observations (see fig.~\ref{fig.atmosp}):  
it starts at $8\times10^5~\mathrm{K}$ at the coronal base, peaks at about
$3\times10^6~\mathrm{K}$ at $3~R_\odot$ and then falls off with distance as
$r^{-0.7}$ \citep{Casalbuoni_al_1999}.
The wind speed profile follows from the wind equations with given temperature
and flux tube expansion, of the form $A(r) = f(r) r^2$, with $f$ a function
which has a maximum close to the coronal base and tends to a finite value at
large distances (see \citealp{Kopp_Holzer_1976}, and
\citealp{Munro_Jackson_1977}). 
The same functional form is chosen for the expansion in the static part of the
atmosphere but different parameters are selected in order to obtain realistic
values for the magnetic field and its continuos variation at the transition
region.
In the photosphere and chromosphere the profile for Alfv\'en speed 
is obtained from the magnetic flux conservation, $B=B_0{A_0}/{A(r)}$,
and the density profile imposed.\\

Following \citet{Dmitruk_al_2001a} we choose the following 
model for the nonlinear terms in eq.~\ref{eq.Alf_wind}
\be
NL^\pm_{j}={\bf z^\pm}(\omega_j)\frac{|Z^\mp|}{L(r)}
\ee\label{NLterm} 
where $L$ represents an integral turbulent dissipation length and
$|Z^\mp|$ stands for the total amplitude of the Els\"{a}sser field
integrated over the whole spectrum ($\Omega$) at the point $r$, hence 
$|Z^\mp|=\sqrt{\int_\Omega [|{\bf z}^\mp(\omega)|^2/\omega~~d\omega]}$.\\
This choice overestimates the transfer rate between high-frequency modes, for which the
Alfv\'en effect is important (shebalin et al?). In reality the predominant interaction, as will be seen 
below, concerns the lowest frequency reflected mode and the full outward propagating spectrum, for which the resonance effects are not important.

The energy distribution among the modes 
influences the dissipation
rate of all the waves coupled. In particular, at a fixed total rms energy,
dissipation is reduced if the energy of the
higher frequency waves is comparable to the lower frequency ones
(flatter spectra) with respect to the
case in which most of the energy is contained in the low frequency
modes (steeper spectra) \citep{Verdini_al_2005}.\\ 
The eqs.~\ref{eq.Alf_wind} can be simplified including the systematic variation
of the Els\"{a}sser amplitude in a new normalized variable
${\bf z}^\pm_N={\bf z}^\pm_O(M_a\pm1)/\sqrt{M_a}$, which reduces to 
${\bf z}^\pm_N=\pm{\bf z}^\pm_O\rho^{1/4}$ in the limit of small
alfv\'enic mach number $M_a=U/V_a\rightarrow0$
(cfr. \citealp{Heinemann_Olbert_1980}).
After Fourier transforming in time the linear equations 
and adding the phenomenological nonlinear term one obtains,
\be\label{eq.Alf_wind_norm} 
(U\pm V_a){\bf z}_N^{\pm'}-i\omega{\bf z}_N^\pm-
\frac{1}{2}(U\pm V_a)\frac{V_a'}{V_a}{\bf z}_N^\mp
=-\frac{|Z_O^\mp|}{L}{\bf z}_N^\pm 
\ee
(the prime indicates a derivative with respect to $r$). The numerically
integrated equations are:
\be\label{eq.Alf_wind_code} 
{\bf z}_N^{\pm'}-i\frac{\omega}{U\pm V_a}{\bf z}_N^\pm-
\frac{1}{2}\frac{V_a'}{V_a}{\bf z}_N^\mp
=-\frac{|Z_O^\mp|}{(U\pm V_a)L}{\bf z}_N^\pm 
\ee 
for the corona, 
while for the photosphere and the chromosphere one gets:
\be\label{eq.Alf_nowind_code}
{\bf z}_N^{\pm'}\mp i\frac{\omega}{V_a}{\bf z}_N^\pm
+\frac{1}{2}\frac{V_a'}{V_a}{\bf z}_N^\mp= 
\mp\frac{|Z_O^\mp|}{V_a L}{\bf z}_N^\pm 
\ee 
The second, third and last coefficient in
eqs.~\ref{eq.Alf_wind_code}-\ref{eq.Alf_nowind_code} represent the
propagation (P), reflection (R) and nonlinear dissipation (NL) coefficients
respectively (inverse of parallel wavelength, reflection scale height, 
nonlinear length scale).
The dissipative feature of the nonlinear terms
can be shown multiplying the above eq.~\ref{eq.Alf_wind_norm}, in its old
variables form, by the
complex conjugate ${\bf z^\pm}^*$ to obtain the evolution equations for 
the Els\"{a}sser energies at a given frequency $E^\pm\equiv\frac{1}{2}|{\bf
  z^\pm}(\omega)|^2$. 
On the RHS one gets $-|{\bf
  z^\pm|^2|Z^\mp|}/L$, which is independent of the phase
difference between the two fields and involves the total amplitude of
the fluctuations (the same term appears in the equation for a static 
atmosphere).
In the presence of a wind, energy flux as conserved quantity is replaced, for
linearly propagating waves, by the total wave action flux, which may be written
as the difference between an outgoing and ingoing flux:
\bd
S^*=S^+-S^-=
\ed
\be
\frac{1}{4}\rho U A 
\left[
\frac{U+V_a}{UV_a}\left({\bf U + V_a}\right)|{\bf z}^+|^2-
\frac{U-V_a}{UV_a}\left({\bf U - V_a}\right)|{\bf z}^-|^2
\right]
\ee\label{eq.wad}
($+,-$ refer to
outward/inward direction and $S$ is the wave action). 

The inward wave action
density vanishes at the Alfv\'en critical point ($\mathrm{X_a}
\approx13~R_\odot$, where the Alfv\'en speed equals the wind speed), so one may write
$S^*=S_0^+-S_0^-=S^+_c$, where the index $c$ stands for the critical point,
while the index $0$ refers to the base of the layer.
Amplitude and the phase of the outward
propagating Els\"{a}sser field (${\bf z}^+$) at $\mathrm{X_a}$ define the natural
boundary conditions, since the critical point is a regular singular point for
the incoming wave equation, because phase velocity of the mode vanishes there:
total wave action density is imposed and the
amplitude and phase of ${\bf z}^-$ can be derived demanding the regularity
of the solutions at $\mathrm{X_a}$.
However, boundary conditions are chosen to assure an
amplitude of the rms velocity field fluctuations 
(i.e. summed over the whole spectrum)
of $\approx40~\mathrm{km~s^{-1}}$ at $1~R_\odot$, as constrained by
observations \citep{Banerjee_al_1998}, with an assigned spectral distribution:
this requires some trial and error since 
nonlinearity does not allow rescaling of the
photospheric amplitude by simply rescaling values at the critical
point $\mathrm{X_a}$.
The shape of photospheric spectrum is imposed approximately 
thanks to the quasi-linear properties of the waves in the 
photosphere-chromosphere layer (small wave amplitudes) and the 
fact that transmission and nonlinearity yield frequency-independent
evolution in the low corona, as shown in the next section.  
Given a slope $p$ at the Alfv\'enic critial point,
the transmission coefficient of the static layer $T(\omega)$ 
(see \citealp{Krogulec_Musielak_1998} for discussion on it),
\be \label{eq.transm}
T(\omega) = \frac{S^+_c}{S^+_0}=
\frac{\rho_c V_{ac}}{\rho_0 V_{a0}}\frac{|{\bf z}_c^+|^2}{|{\bf z}_0^+|^2}=
\frac{|{\bf z}_{Nc}^+|^2}{|{\bf z}_{N0}^+|^2}
\ee
can therefore be 
used to correct the initial spectrum 
$|{\bf z}^+(\omega)|=|{\bf z}^+(\omega_0)|\times(\omega/\omega_0)^p$
to the desired spectrum at the photosphere imposing
$|{\bf z}^+(\omega)|=|{\bf z}^+(\omega_0)|\sqrt{T(\omega)}\times(\omega/\omega_0)^p$.
In order to describe the spectrum 32 modes are chosen in the range of
frequency between $10^{-6}~\mathrm{Hz}$ and $10^{-2}~\mathrm{Hz}$ with
increasing resolution at higher frequencies.\\

The phenomenological turbulent length scale varies as
$L(r)=L_0\times\sqrt{A(r)}$, where $L_0=34,000~\mathrm{km}$ 
is imposed at the coronal base and corresponds to the average size of the 
supergranules.
The waves are propagated from the Alfv\'enic critical point
forward (to the Earth orbit) and backward (till the base of
the corona) by integration of eqs.~\ref{eq.Alf_wind_code}.
The conservation of the energy flux across the transition region allows one to
determine the Els\"{a}sser fields below the discontinuity 
which are propagated back to the base of the photosphere 
using eqs.~\ref{eq.Alf_nowind_code}.

\section{RESULTS}
\clearpage
\begin{figure*}[th]
\plotone{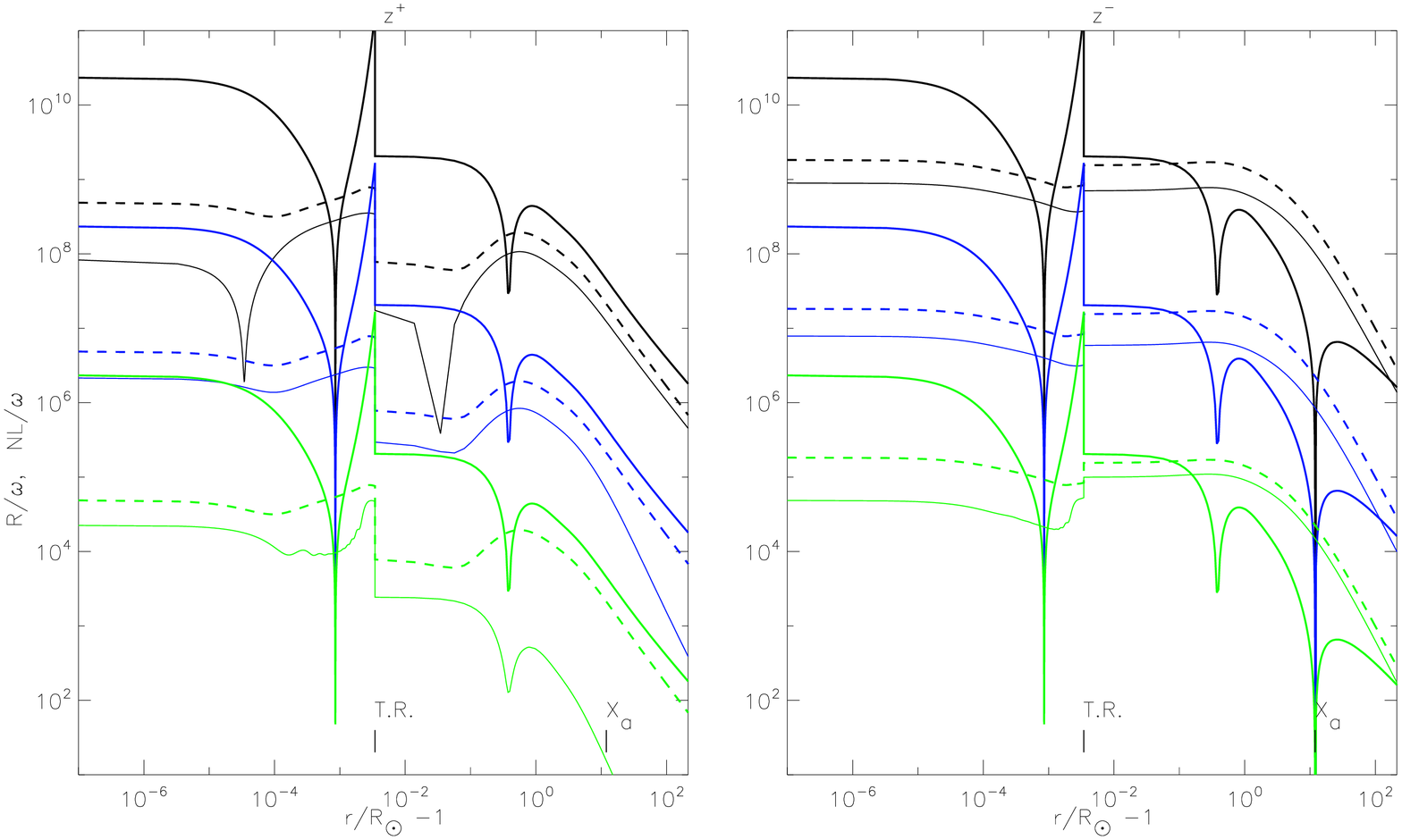}
\caption{
Comparison of the reflection (R, thick solid line) and nonlinear coefficient
(NL, thick dashed line) normalized to the
propagation coefficient (P) for the outgoing and ingoing wave at three different
frequencies ($10^{-6}~\mathrm{Hz},~10^{-4}~\mathrm{Hz},~10^{-2}~\mathrm{Hz}$;
black, blue and green lines respectively).
Also shown in thin solid line is the contribution of each frequency wave to the
nonlinear coefficient. The photospheric frequency spectrum is flat and the
boundary conditions at $\mathrm{X_a}$ are set to get a rms velocity field fluctuation 
at the coronal base $\delta u=40~\mathrm{km~s^{-1}}$.
}
\label{fig.scaleh_flat}
\end{figure*}
\clearpage
Following \citealp{Velli_1993} we compare the characteristic lengthscales
of eqs.~\ref{eq.Alf_wind_code}-\ref{eq.Alf_nowind_code} in the two layers. 
First consider the thick lines in fig.~\ref{fig.scaleh_flat} which represent
the reflection and nonlinear coefficients
(solid and dashed line respectively) normalized to the propagation coefficient
for $\omega=10^{-6}~\mathrm{Hz},~10^{-4}~\mathrm{Hz},~10^{-2}~\mathrm{Hz}$
(black, blue and green lines respectively)
for a flat photospheric spectrum. 
Reflection has a maximum at the transition region and it falls off 
by a factor of about 100 in the corona (because of the density drop).
The zeros in the reflection coefficient appearing for both the $z^+$ and the
$z^-$ depend on the fact that $V_a'=0$ (approximately in the corona), 
while the one located at $\mathrm{X_a}$ 
appears only for the backward propagating waves since the propagation
coefficient becomes infinite there (see eq.~\ref{eq.Alf_wind_code}).\\
For the outward propagating wave (left panel) reflection is generally much 
greater (a factor 100) than dissipation in the photosphere-chromosphere 
and in the very low corona (below $\approx1.2R_\odot$).
Further out the nonlinear dissipation is smaller than reflection but of the same
order of magnitude.
For the inward propagating wave (right panel), again reflection dominates in the
photosphere-chromosphere (by a factor of 10), 
but in the corona the dissipative coefficient is
comparable or much greater than the reflection coefficient.

\begin{figure}[th]
\plotone{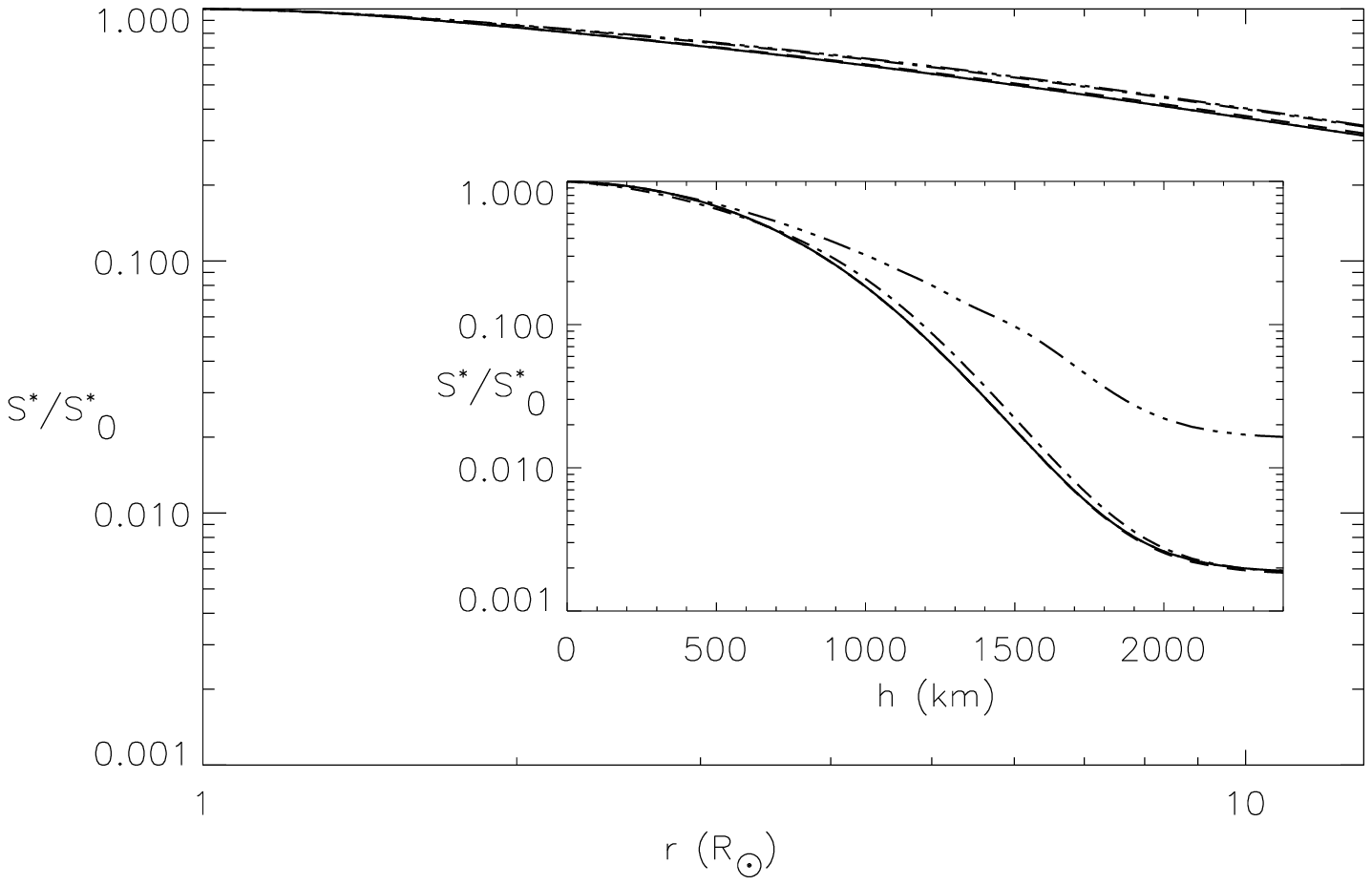}
\caption{
Normalized wave action density for
the corona as function of distance
for 5 frequencies 
($10^{-6}~\mathrm{Hz}$ solid line,
$10^{-5}~\mathrm{Hz}$ dotted line,
$10^{-4}~\mathrm{Hz}$ dashed line,
$10^{-3}~\mathrm{Hz}$ dotted-dashed line,
$10^{-2}~\mathrm{Hz}$ triple-dotted-dashed line).
The wave energy flux for the photosphere-chromosphere is plotted in the
subpanel whit the same line coding.}
\label{fig.Diss_rel}
\end{figure}

\begin{figure}[th]
\plotone{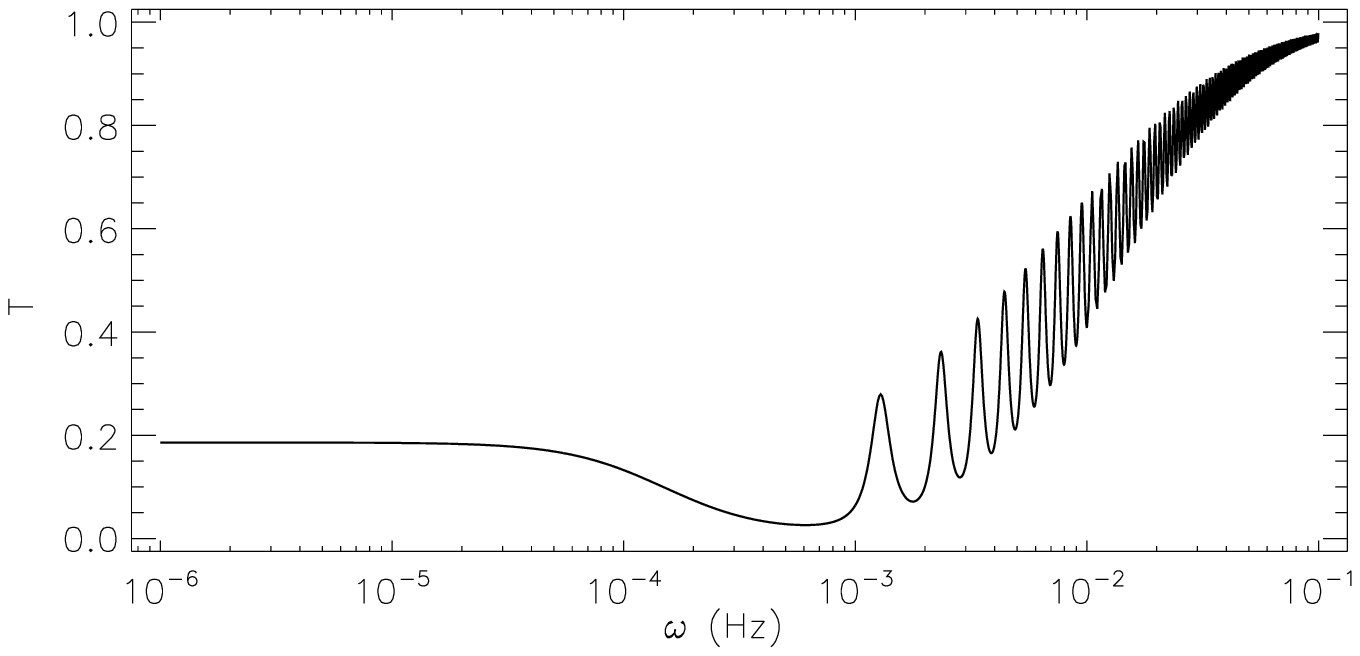}
\caption{Transmission coefficient for the photosphere-chromosphere 
as function of frequency.
}
\label{fig.Trans_phot}
\end{figure}

The \emph{relative} dissipation
of the linearly conserved quantities,
as defined below in eq.~\ref{eq.rel_diss},
has hence different features in the two layers.
In fig.~\ref{fig.Diss_rel} we plot   
the total wave action density for the corona (main panel) 
and the total wave energy flux for the static layer (sub panel) 
normalized to their base value for all the frequencies which form the
spectrum, i.e. 
\bd
\frac{S^*(r,\omega)}{S^*_0(\omega)}=
\frac{|{\bf z}^+_N|^2-|{\bf z}^-_N|^2}{|{\bf z}^+_{N_0}|^2-|{\bf z}^-_{N_0}|^2}=
\ed
\be\label{eq.rel_diss}
1-\frac{1}{2\left(|{\bf z}^+_{N_0}|^2-|{\bf z}^-_{N_0}|^2\right)}
\int_{r_0}^r\frac{\mathrm{d}r}{LV_a} 
            \left(\frac{|Z^-_O|}{1+M_a}|{\bf z}^+_N|^2 + 
                  \frac{|Z^+_O|}{1-M_a}|{\bf z}^-_N|^2\right), 
\ee
with the normalization used to derive
eqs.~\ref{eq.Alf_wind_norm}:
the coefficients appearing in the integral are the nonlinear
frequency integrated coefficients discussed above.\\
\clearpage
\begin{figure}[th]
\plotone{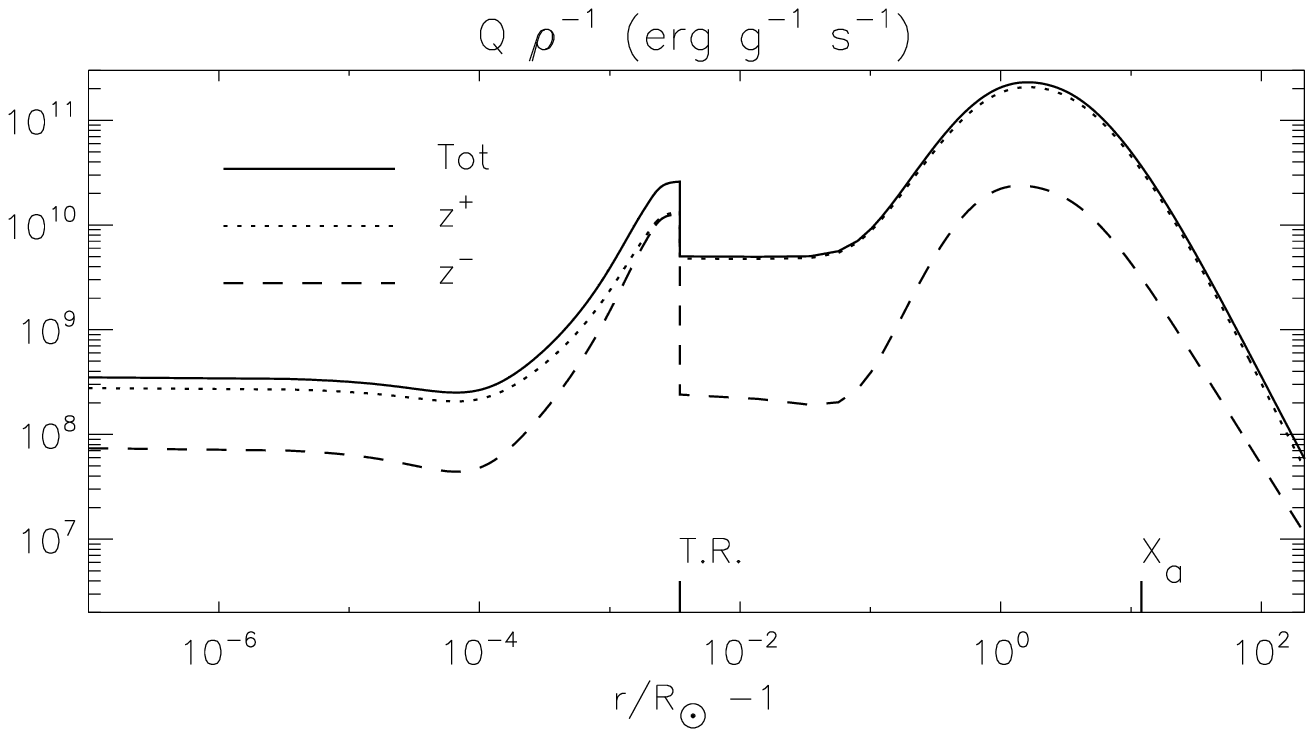}
\caption{
Heating rate per unit mass integrated over the whole spectrum as function of
heliocentric distance. The contribution of the ingoing (dashed line) and
outgoing (dotted line) heating
rate is also shown. The transition region (T.R.) and the Alfv\'enic critical
point ($\mathrm{X_a}\approx13~R_\odot$) are indicated on the x axis.
}
\label{fig.Heat_mass}
\end{figure}
\clearpage
In the upper chromosphere the flux tube expansion is very rapid and reflection
is strong,
both the ingoing and outgoing wave contribute to the damping of the energy flux
(comparable -less than one order of magnitude difference- 
nonlinear coefficient and wave amplitudes) 
and the relative dissipation is very high.
Low frequency modes (lower plot in the subfigure) are 
the most damped (the most reflected)  
while high frequency modes (higher plots) are the less damped.
In fact, inspection of eq.~\ref{eq.rel_diss} reveals that the relative 
dissipation is quadratic in the frequency dependent wave amplitudes 
($|{\bf z}_N^\pm|^2$) which in turn increase with decreasing frequency because
of the different reflection rate.
In the corona, instead, beyond $2R_\odot$ 
the dissipation coefficient for the outgoing waves is weaker
and their amplitudes grow,
reflection is weaker as well, and an imbalance
between outgoing and ingoing fluxes holds. 
Only the former
contribute to the wave action density dissipation since now 
the dominant quadratic dependence in eq.~\ref{eq.rel_diss} 
comes form the outgoing mode 
(see for example the approximate conservation form used by
\citealp{Cranmer_Ballegooijen_2005}).
Note that for all frequencies
the wave action density decreases at approximately the same rate.
It turns out that 
the amplitude evolution is driven mainly by the nonlinear, frequency
independent, term in the corona and by the reflection, frequency dependent, 
term in the photosphere-chromosphere, a feature we will find again studying
the power spectrum evolution.\\
In comparison the heating rate per unit mass,
an \emph{absolute} measure of energy dissipation,
integrated over the spectrum,
\be\label{eq.Heat}
\frac{Q}{\rho}=\frac{Q^+}{\rho}+\frac{Q^-}{\rho}=\frac{
|Z^+|^2|Z^-|+|Z^-|^2|Z^+|}{L(r)},
\ee
is generally higher in the corona than in the photosphere-chromosphere,
as shown in fig.~\ref{fig.Heat_mass}.
In the latter layer both
the ingoing and outgoing wave contributes to the total amount of heating rate,
while in the former most of the dissipation comes from the outgoing mode.
The absolute dissipation is quadratic in the frequency integrated 
wave amplitudes and in the corona
outgoing wave are allowed to grow almost undamped (low relative dissipation)
but the existence of a small seed of ingoing wave assures
a large absolute dissipation. This is not true in the photosphere-chromosphere,
before the rapid expansion of the flux tube, 
where the wave amplitude is small and there is 
a small imbalance between outgoing
and ingoing propagating wave amplitudes.\\
\clearpage
\begin{figure*}[th]
\plotone{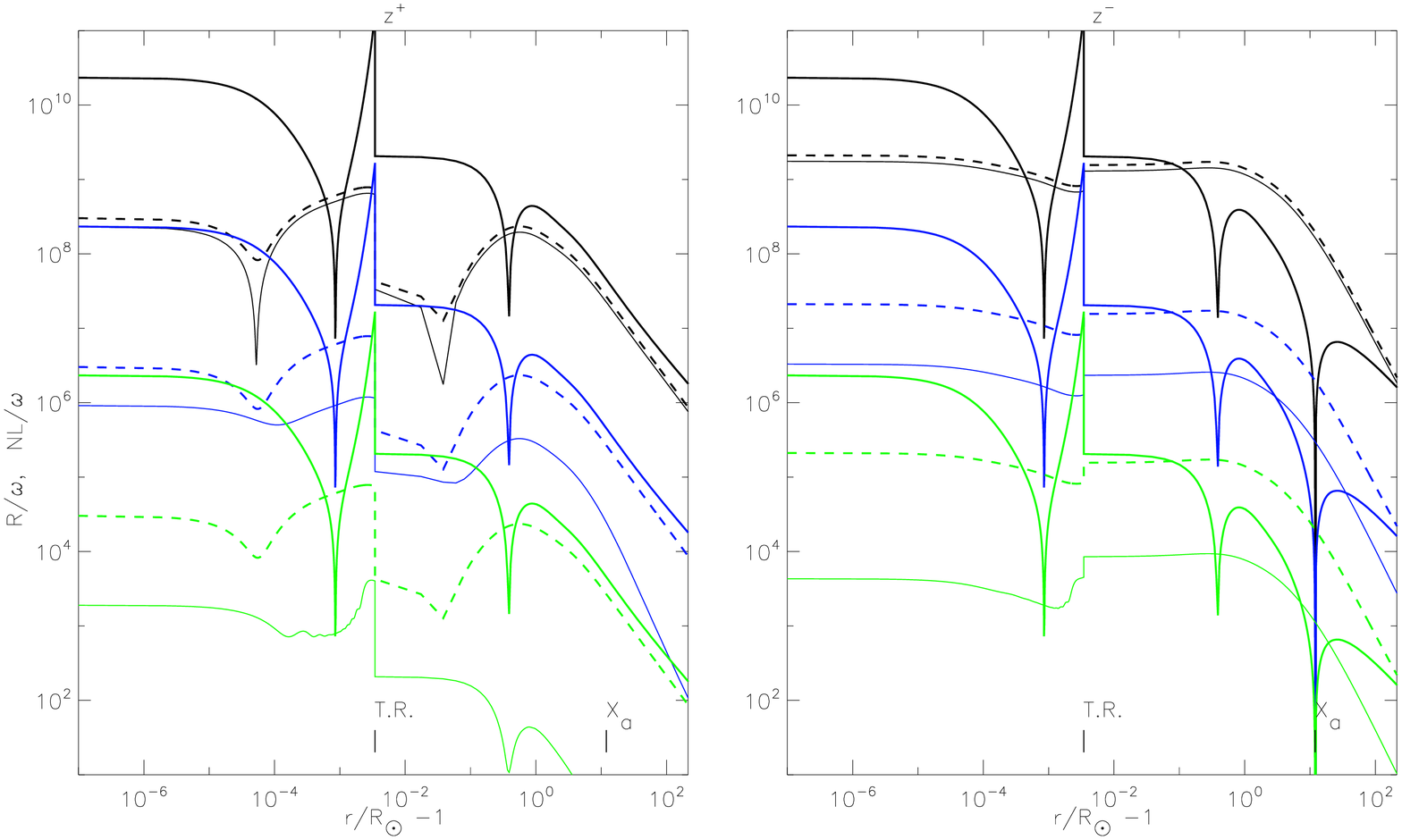}
\caption{
Same as fig.~\ref{fig.scaleh_flat} for a photospheric Kolmogorov-like frequency spectrum}
\label{fig.scaleh_kolm}
\end{figure*}
\clearpage
The effect of a different slope of the initial spectrum can be understood
analyzing the contribution of each frequency to the nonlinear coefficient,
plotted in thin lines in fig.~\ref{fig.scaleh_flat}. 
Starting with a flat photospheric frequency spectrum results in an
approximately equal
contribution to the total nonlinear term in the whole atmosphere, except for
the outer corona where the nonlinear coefficient for the outward propagating
wave is made up of essentially backward propagating
waves at low frequencies.
Note also that the frequency decomposed nonlinear coefficient is
approximately the same 
for outgoing and ingoing propagating waves in the photosphere-chromosphere
since reflection is high enough compared to dissipation.
It follows that if a Kolmogorov-like photospheric spectrum 
($E/\omega\propto\omega^{-5/3}$) is imposed, the nonlinear term is
mainly made up of low frequency waves 
for both counterpropagating waves, in both the layers.
This is can be seen in fig.~\ref{fig.scaleh_kolm} comparing the dashed thick
lines and the solid thin lines: 
for $\omega\age 10^{-4}~\mathrm{Hz}$ the contribution to the nonlinear 
coefficient is generally less then $10\%$.
An exception is found below $2R_\odot$ for the outgoing mode, since
reflection is high even for intermediate frequency wave 
(see fig.~\ref{fig.Trans_phot} for the photospheric layer).
Note that a deap in the (frequency integrated) nonlinear coefficient 
for the outgoing mode appears below the location of vanishing reflection 
in both the photosphere and low corona, 
since the energy resides mainly in the low frequency mode.\\

\clearpage
\begin{figure}[th]
\plotone{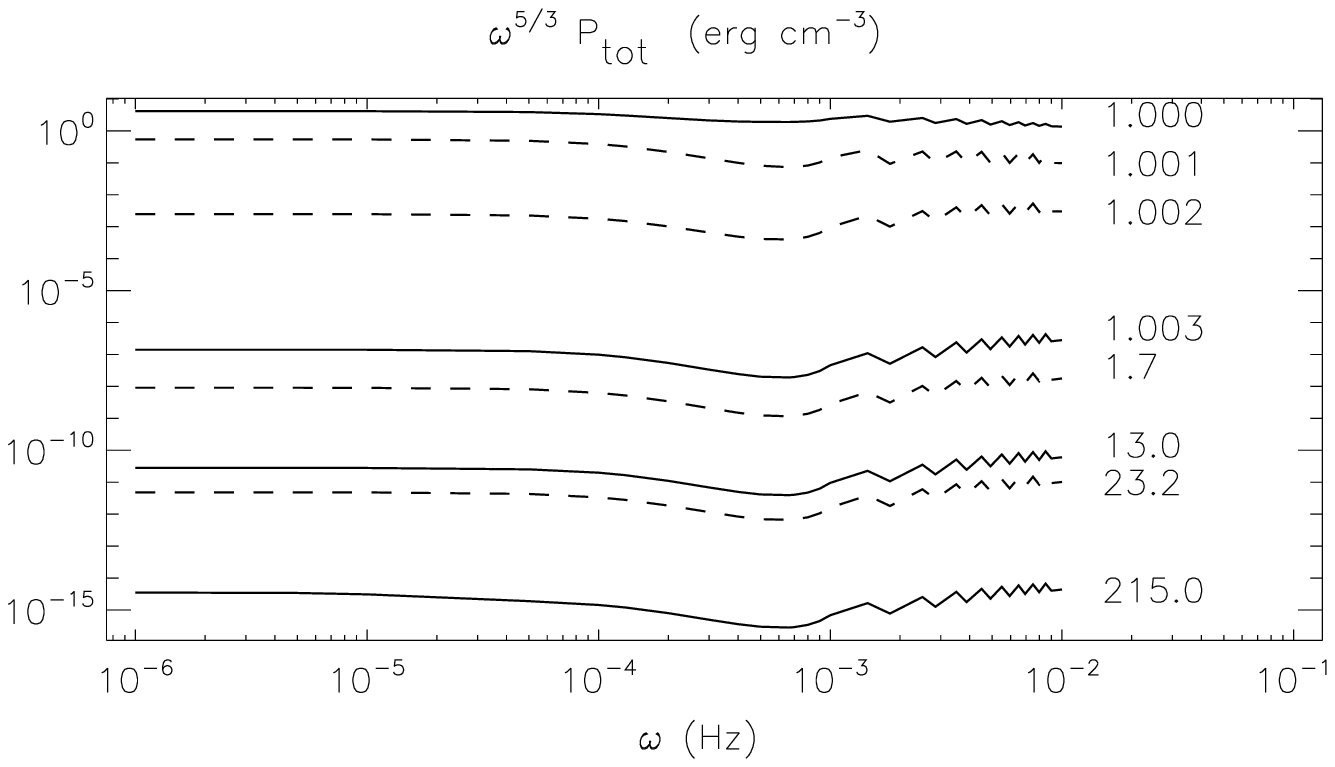}
\caption{Compensated power spectrum as function of heliocentric distance
for a photospheric kolmogorv-like initial spectrum. Each curve is labelled with
the corresponding heliocentric distance in unit of $R_\odot$.
From top to bottom, solid lines indicate the photospheric base, 
the T.R., $\mathrm{X_a}$ and 1~AU. 
}
\label{fig.Spectrum}
\end{figure}
\clearpage
This separate behavior in the two layers has strong consequences on
spectral evolution. In fig.~\ref{fig.Spectrum} the (compensated) total 
power in the fluctuations is plotted for different heliocentric distances.
An almost Kolmogorow-like spectrum is imposed at the base of the photosphere
with a procedure described at the end of section~\ref{sec.model}.
At very low frequencies the spectrum practically does not evolve, in the whole 
domain, while there is a tendency to steepen at low-intermediate frequencies
($10^{-5}~\mathrm{Hz}\ale\omega\ale10^{-3}~\mathrm{Hz}$).
The behavior at high frequencies is quite complicated. Some irregularities
appear very close to the base of the photosphere and the overall tendency is
that of flattening. Note, however, that most of the changes in the shape 
occur in the photosphere-chromosphere 
where the waves display a strong
frequency dependent behavior. 
This makes the spectral evolution very similar to
the linear case, below the transition region (except the energy level of the
spectrum), and the appearance of the irregularities can be interpreted by
means of the linear analysis. 
Accordingly, in fig.~\ref{fig.Trans_phot} we plot the transmission coefficient,
defined in eq.~\ref{eq.transm}, as function of frequency for 
the photosphere-chromosphere.  
Note that the transmission is constant at low frequency, decreases at
intermediate frequencies and increase again at high frequencies, where
several transmission peaks appear: basically all spectral evolution is
qualitatively reproduced.\\
The peaks originate form the discontinuity in
the reflection scale height at the transition region \citep{Velli_1993}. In
fact the amplitude of the reflected waves shows some nodes inside the domain
and when their location coincides with the base of the photosphere the
transmission is enhanced (a condition which depends on the frequency of the
waves, see \citealp{Hollweg_1978a}).
When nonlinearities are introduced the location of the nodes depends also on
the wave amplitude imposed at $\mathrm{X_a}$ (see \citealp{Verdini_al_2005})
and similarly if these nodes 
are located near the base of the photosphere 
the irregularities in the spectrum appear.\\ 

\begin{figure}[th]
\plotone{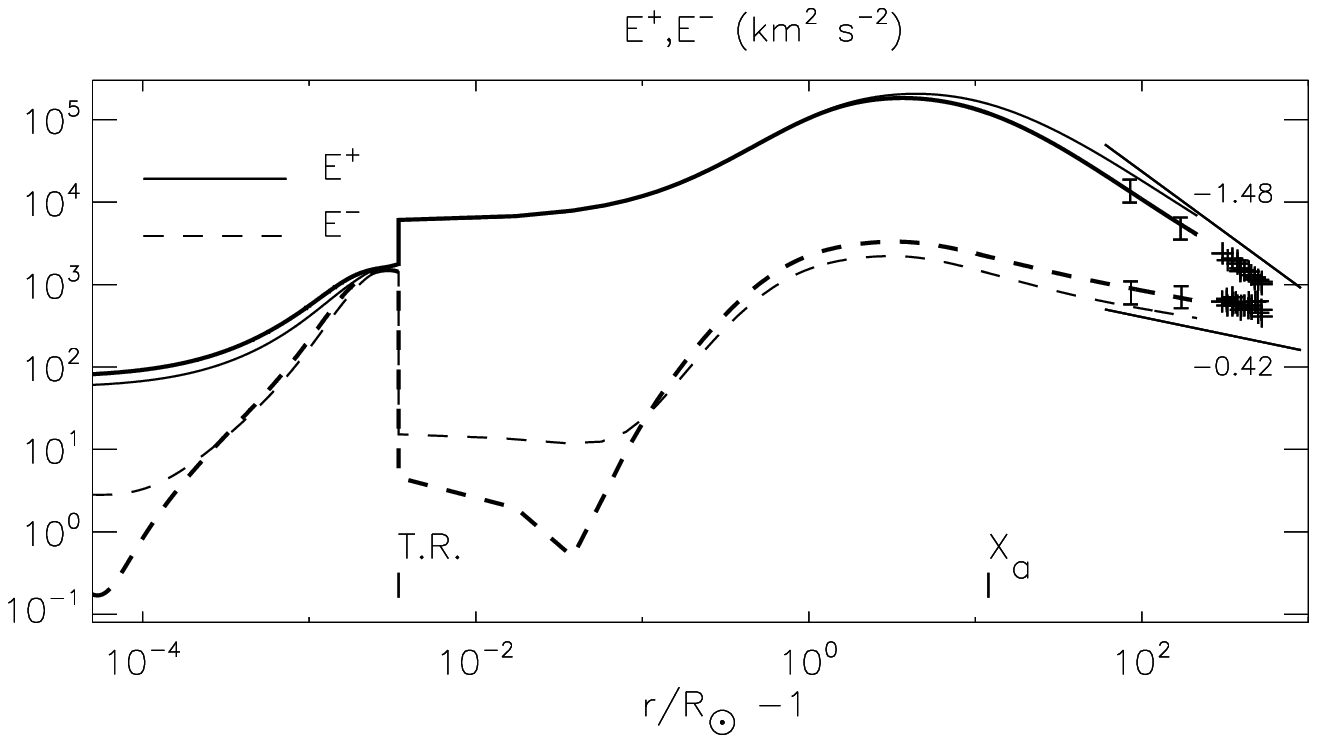}
\caption{Frequency integrated Els\"{a}sser energies
as function of heliocentric distance for a photospheric Kolmogorov
spectrum with $\delta u=40~\mathrm{km~s^{-1}}$ at the coronal base. Symbols
indicate observational constraints (see text for explanation).}
\label{fig.elsen_kolmflat}
\end{figure}

\begin{figure*}[th]
\plotone{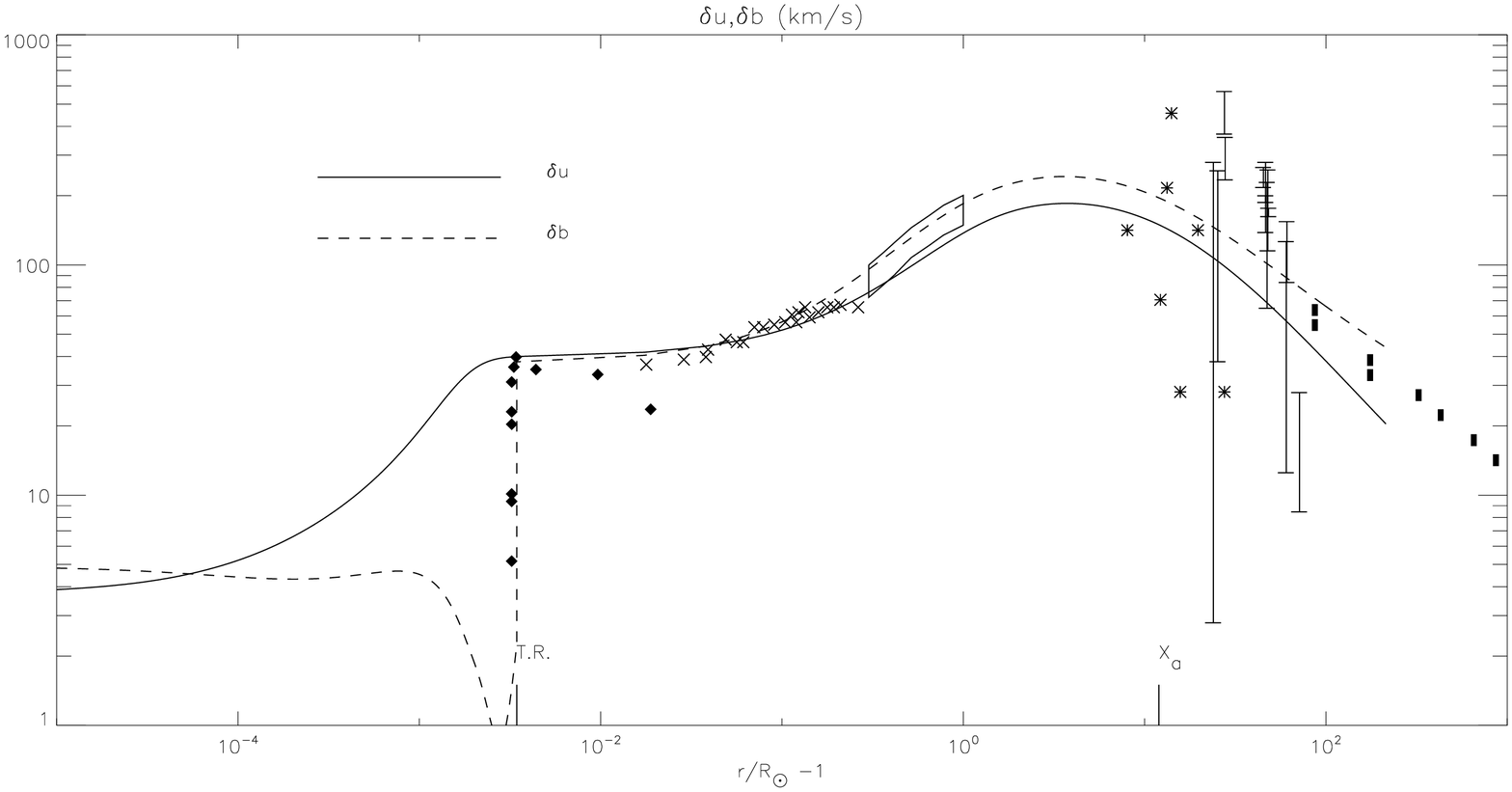}
\caption{Root mean square amplitude $\delta u$ and $\delta b$ (in velocity
units) as function of heliocentric distance for a photospheric Kolmogorov
spectrum with $\delta u=40~\mathrm{km~s^{-1}}$ at the coronal base. Symbols
indicates observational constraints (see text for explanation).}
\label{fig.rmsamp_kolm}
\end{figure*}

The slope of the spectrum imposed at the photosphere has negligible effects on
the total power spectral evolution, however it changes the amount of
energy residing in the ingoing and outgoing mode (or in the kinetic or magnetic
fluctuations) at large distances 
and some constraints on the slope can be obtained using the
available observational data. 
In fig.~\ref{fig.elsen_kolmflat} 
the Els\"{a}sser energies $E^\pm$ integrated over
the frequency spectrum are plotted (solid and dashed line respectively)
along with the Ulysses and Helios data \citep{Bavassano_al_2000a}, for a 
Kolmogorov (thick lines) and 
a flat (thin lines) initial slope with 
$\delta u=40~\mathrm{km~s^{-1}}$ at the coronal base.
Both the data and the expected slopes are reproduced by the Kolmogorov-like
photospheric spectrum while the flat one has to high outgoing energy and too
low ingoing energy. The effect of high energy at high frequency waves
is that of dissipating the inward wave,
since they are mainly outward propagating: as a result outgoing waves are
allowed to propagate almost undamped and their energy content is hence higher.
Note that in the Kolmogorov case a deap, very close to the coronal base, appears
as a signature of vanishing ingoing waves, a feature of the low frequency
reflected waves.
This results in a vanishing absolute dissipation (heating) 
which is not found for the flat case and has important consequences for the
acceleration and heating of the solar wind.\\
\clearpage
\clearpage
In the following we consider only a Kolmogorov spectrum.
In fig.~\ref{fig.rmsamp_kolm} the root mean square amplitude of velocity field
fluctuation integrated over the whole spectrum is plotted as function of
heliocentric distance (solid line, in dotted line we plot also the magnetic
field fluctuation in velocity unit) along with some observational
data (taken from \citealp{Cranmer_Ballegooijen_2005}, to which
we address for comment on the data set):\\
\begin{itemize}
\item Filled diamonds are nonthermal line broadening velocities measured by
SUMER on the disk \citep{Wilhelm_al_1995},
\item Crosses are nonthermal velocities derived from SUMER
observations above the solar limb \citep{Banerjee_al_1998}
\item The box represents the upper and lower limit given by
\citealp{Esser_al_1999} from UVCS off-limb data
\item Stars are early measurements from \citealp{Armstrong_Woo_1981}
\item The bars are recent measurements of transverse velocity field
fluctuation using radio scintillation \citep{Canals_al_2002} 
\item Filled bars are the Helios and Ulysses data for the Els\"{a}sser
energies, from \citealp{Bavassano_al_2000a}, 
rewritten in term of the velocity field
fluctuation assuming equipartition between magnetic and kinetic energy.
\end{itemize} 
Note that the Helios and Ulysses data are obtained averaging over periods
$\age$1~hour (correspondig to $\omega\ale10^{-4}~\mathrm{Hz}$) 
while all the other points in the figure refers to rms values.
\clearpage
\clearpage
The overall agreement is quite good, even if data suggest
a smaller power (more dissipation) just above the T.R. and more power (less
dissipation) at about $2R_\odot$.
Note that because of the equipartition assumption the Helios and Ulysses
data disagree with the integrated quantities (the correct comparison has
already been made above in fig.~\ref{fig.elsen_kolmflat}).
As noted by \citealp{Cranmer_Ballegooijen_2005} the longitudinal velocity
fluctuation data (filled diamonds) agree very well with the magnetic field
fluctuation amplitudes (dashed line) and could indicate wave coupling among
transverse and longitudinal mode. At leading order compressive effects are
driven by the magnetic pressure originating from the incompressible fluctuations
and represent a way for Alfv\'en waves to get rid of the energy excess above the
T.R.. If these compressional waves are isotropic and suffer some dumping via
shock formation, or other processes active in the low corona, 
they can reproduced the measured parallel $\delta u$ and supply the heating
needed by current model of wind acceleration.\\

\section{CONCLUSIONS}
In this paper we have modeled the nonlinear evolution of Alfv\'en waves
propagating through the photosphere, the corona and the solar wind till 1~AU.
Nonlinear interactions occur between outward propagating and
reflected waves, and it is assumed that a nonlinear cascade develops 
preferentially in a direction perpendicular to that of propagation, 
which we take to coincide with the direction of
the mean radial magnetic field.\\
While the phenomenological nonlinear term acts as a
dissipative sink for both outward and inward waves,
independently of the wave frequency, 
reflection, provided by the stratification of the layer, is 
generally strong at low frequencies and decreases with increasing frequency.\\

We find that 
most of the heating occurs in the low corona (below the Alfv\'enic
critical point), while very little power is dissipated below the 
transition region.
For reasonable velocity field fluctuations at the base
of the photosphere a sufficient amount of energy flux is transmitted through
the transition region.
The adopted frequency coupling is not able to reproduce the observed
spectral slope and evolution in
the Alfv\'enic range even though frequency integrated data at large distances 
constrain the outer spectrum to be steep (-5/3 slope).
The modification of the frequency spectrum occurs mainly in the
chromosphere and in the photosphere since waves experience a 
strong reflection at all the frequencies considered, 
while in the corona and the solar wind the spectrum maintains approximately 
the same shape one finds at the coronal base.\\

Nonlinear dissipation based on reflection acts in 
different ways depending
both on the (ingoing and outgoing) wave amplitude  
and on the layer considered.
In the corona, reflection is not very high but the
outgoing wave amplitude is allowed to grow, so that 
the wave evolution is driven by the nonlinear interactions
(all the modes evolve in the same way)
and one finds a strong heating rate in the Sub-Alfv\'enic corona.\\
In the photosphere-chromosphere a strong reflection rate, combined with 
small wave amplitudes, leads to an evolution similar to the linear case, 
which depends on frequency, and a small heating rate.\\
As a result most of the wave energy dissipation takes place in the 
first 4 solar radii above the coronal base.
The driving modes for dissipation are the modes
which experience the biggest reflection, generally low frequency modes.
However depending on the model of atmosphere, i.e. on its characteristic scale
height, and on the energy distribution, i.e. flat or steep spectra,
intermediate frequency modes can be important as well.\\

The spectral shape varies mainly below the transition region, 
it steepens at low-intermediate frequencies
($10^{-5}~\mathrm{Hz}\ale\omega\ale10^{-3}~\mathrm{Hz}$) 
and flattens at high
frequencies ($\omega\age10^{-3}~\mathrm{Hz}$) showing the
characteristic features (energy peaks and frequency distribution) 
one finds in the transmission coefficient (linear behavior).
In the corona instead, it maintains
approximately the shape one finds beyond the T.R.,
because of the form of the nonlinear term adopted. 
The very low frequency range
($\omega\ale10^{-5}~\mathrm{Hz}$) 
practically does not evolve in
the whole layer and it keeps the original slope at the photosphere. 
With this model of nonlinearities one can conclude that
the spectrum one finds at 1 AU is basically the same spectrum 
at the base of the corona.\\
The input spectrum at the photosphere, whatever
the shape is, is instead strongly modified by the 
transmission properties of the
atmosphere below the transition region 
(independently of the model used for the nonlinear interaction).
The energy peaks in the spectrum, resulting from an enhanced transmission 
at high frequencies, indicate that, even in presence of nonlinar interactions,
the photospheric layer act as a filter 
for the energy injected through photospheric footpoint motion, if a smoothing
of the forcing frequency is to be present, it must occur in this highly
stratified layer.\\
The data at large distances suggest that energy at high frequency 
should be very low,
however we find an energy increase at high frequency.
Since the spectral evolution in corona depends also on the approximate
frequency coupling contained in the nonlinear term, constraints on 
the photospheric input spectrum can not be given safely. 
Given that high frequency waves are transmitted through the T.R. and are quite
energetic in the very low corona
some other mechanism must be invoked to dissipate high frequency waves 
or a better modeling of the nonlinearities, which we plan to do in future
works.\\
As first pointed out by \citealp{Hollweg_1981}, such high frequency energy
reservoir can be the source for plasma heating processes 
operating in the low corona. 
Note that not only the peaks contribute to the
energy budget, but the general flattening of the spectrum is 
important as well.\\
A comparison with measurements of $\delta u$ suggests
that the model can be considered a very good approximation in the outer
corona and solar wind while, despite the good agreement  found in the
low corona, some other processes must be invoked to reproduced the
observed features below the alfv\'enic critical point, 
such as compressible effects and wave coupling, especially in the chromosphere
and photosphere. 
Other models of turbulent transport have been constructed to fit the decay of
turbulence with distance from the sun in the solar wind beyond 1 AU
(\citealp{Smith_al_2001,Breech_al_2005})  as well as to explain the extended heating in this region. Here the Alfv\'en speed can be 
neglected in the transport of the fluctuations, so that in some sense our model equations should be consistent with theirs, when rewritten in terms of the second order moments. A generalisation of turbulence transport equations, consistent both in the corona,  acceleration region and solar wind is a topic of current research.

\acknowledgements

We would like to thank the IPAM program ``Grand Challenge Problems in
Computational Astrophysics'' at UCLA where this work was completed.
We also thank S. Oughton and W. H. Matthaeus for useful discussion.\\


\begin{thebibliography}{46}
\expandafter\ifx\csname natexlab\endcsname\relax\def\natexlab#1{#1}\fi

\bibitem[{{Armstrong} \& {Woo}(1981)}]{Armstrong_Woo_1981}
{Armstrong}, J.~W. \& {Woo}, R. 1981, Astron.\ Astrophys., {\bf 103}, 415

\bibitem[{{Banerjee} {et~al.}(1998){Banerjee}, {Teriaca}, {Doyle}, \&
  {Wilhelm}}]{Banerjee_al_1998}
{Banerjee}, D., {Teriaca}, L., {Doyle}, J.~G., \& {Wilhelm}, K. 1998, Astron.\
  Astrophys., {\bf 339}, 208

\bibitem[{{Bavassano} {et~al.}(2000{\natexlab{a}}){Bavassano}, {Pietropaolo},
  \& {Bruno}}]{Bavassano_al_2000b}
{Bavassano}, B., {Pietropaolo}, E., \& {Bruno}, R. 2000{\natexlab{a}}, J.\
  Geophys.\ Res., {\bf 105}, 12697

\bibitem[{{Bavassano} {et~al.}(2000{\natexlab{b}}){Bavassano}, {Pietropaolo},
  \& {Bruno}}]{Bavassano_al_2000a}
{Bavassano}, B., {Pietropaolo}, E., \& {Bruno}, R. 2000{\natexlab{b}}, J.\
  Geophys.\ Res., {\bf 105}, 15959

\bibitem[{{Breech} {et~al.}(2005){Breech}, {Matthaeus}, {Minnie}, {Oughton},
  {Parhi}, {Bieber}, \& {Bavassano}}]{Breech_al_2005}
{Breech}, B., {Matthaeus}, W.~H., {Minnie}, J., {et~al.} 2005, Geophys.\ Rev.\
  Lett., {\bf 32}, 6103

\bibitem[{{Canals} {et~al.}(2002){Canals}, {Breen}, {Ofman}, {Moran}, \&
  {Fallows}}]{Canals_al_2002}
{Canals}, A., {Breen}, A.~R., {Ofman}, L., {Moran}, P.~J., \& {Fallows}, R.~A.
  2002, Annales Geophysicae, {\bf 20}, 1265

\bibitem[{{Casalbuoni} {et~al.}(1999){Casalbuoni}, {Del Zanna}, {Habbal}, \&
  {Velli}}]{Casalbuoni_al_1999}
{Casalbuoni}, S., {Del Zanna}, L., {Habbal}, S.~R., \& {Velli}, M. 1999, J.\
  Geophys.\ Res., {\bf 104}, 9947

\bibitem[{{Cranmer} \& {van Ballegooijen}(2005)}]{Cranmer_Ballegooijen_2005}
{Cranmer}, S.~R. \& {van Ballegooijen}, A.~A. 2005, ApJS, {\bf 156}, 265

\bibitem[{{Del Zanna} {et~al.}(2001){Del Zanna}, {Velli}, \&
  {Londrillo}}]{DelZanna_al_2001}
{Del Zanna}, L., {Velli}, M., \& {Londrillo}, P. 2001, Astron.\ Astrophys.,
  {\bf 367}, 705

\bibitem[{{Dmitruk} \& {Matthaeus}(2003)}]{Dmitruk_Matthaeus_2003}
{Dmitruk}, P. \& {Matthaeus}, W.~H. 2003, Astrophys.\ J., {\bf 597}, 1097

\bibitem[{{Dmitruk} {et~al.}(2004){Dmitruk}, {Matthaeus}, \&
  {Lanzerotti}}]{Dmitruk_al_2004}
{Dmitruk}, P., {Matthaeus}, W.~H., \& {Lanzerotti}, L.~J. 2004, Geophys.\ Rev.\
  Lett., {\bf 31}, 21805

\bibitem[{{Dmitruk} {et~al.}(2001{\natexlab{a}}){Dmitruk}, {Matthaeus},
  {Milano}, \& {Oughton}}]{Dmitruk_al_2001b}
{Dmitruk}, P., {Matthaeus}, W.~H., {Milano}, L.~J., \& {Oughton}, S.
  2001{\natexlab{a}}, Physics of Plasmas, {\bf 8}, 2377

\bibitem[{{Dmitruk} {et~al.}(2002){Dmitruk}, {Matthaeus}, {Milano}, {Oughton},
  {Zank}, \& {Mullan}}]{Dmitruk_al_2002}
{Dmitruk}, P., {Matthaeus}, W.~H., {Milano}, L.~J., {et~al.} 2002, Astrophys.\
  J., {\bf 575}, 571

\bibitem[{{Dmitruk} {et~al.}(2001{\natexlab{b}}){Dmitruk}, {Milano}, \&
  {Matthaeus}}]{Dmitruk_al_2001a}
{Dmitruk}, P., {Milano}, L.~J., \& {Matthaeus}, W.~H. 2001{\natexlab{b}},
  Astrophys.\ J., {\bf 548}, 482

\bibitem[{{Esser} {et~al.}(1999){Esser}, {Fineschi}, {Dobrzycka}, {Habbal},
  {Edgar}, {Raymond}, {Kohl}, \& {Guhathakurta}}]{Esser_al_1999}
{Esser}, R., {Fineschi}, S., {Dobrzycka}, D., {et~al.} 1999, ApJL, {\bf 510},
  L63

\bibitem[{Habbal {et~al.}(1995)Habbal, Esser, Guhathakurta, \&
  Fisher}]{Habbal_al_1995}
Habbal, S.~R., Esser, R., Guhathakurta, M., \& Fisher, R.~R. 1995, Geophys.\
  Rev.\ Lett., {\bf 22}, 1465

\bibitem[{{Heinemann} \& {Olbert}(1980)}]{Heinemann_Olbert_1980}
{Heinemann}, M. \& {Olbert}, S. 1980, J.\ Geophys.\ Res., {\bf 85}, 1311

\bibitem[{{Hollweg}(1978)}]{Hollweg_1978a}
{Hollweg}, J.~V. 1978, Solar Phys., {\bf 56}, 305

\bibitem[{{Hollweg}(1981)}]{Hollweg_1981}
{Hollweg}, J.~V. 1981, Solar Phys., {\bf 70}, 25

\bibitem[{{Hollweg} {et~al.}(1982){Hollweg}, {Jackson}, \&
  {Galloway}}]{Hollweg_al_1982a}
{Hollweg}, J.~V., {Jackson}, S., \& {Galloway}, D. 1982, Solar Phys., {\bf 75},
  35

\bibitem[{{Kopp} \& {Holzer}(1976)}]{Kopp_Holzer_1976}
{Kopp}, R.~A. \& {Holzer}, T.~E. 1976, Solar Phys., {\bf 49}, 43

\bibitem[{{Krogulec} \& {Musielak}(1998)}]{Krogulec_Musielak_1998}
{Krogulec}, M. \& {Musielak}, Z.~E. 1998, Acta Astronomica, {\bf 48}, 77

\bibitem[{{Leroy}(1980)}]{Leroy_1980}
{Leroy}, B. 1980, Astron.\ Astrophys., {\bf 91}, 136

\bibitem[{Li {et~al.}(1999)Li, Habbal, Hollweg, \& Esser}]{Li_al_1999}
Li, X., Habbal, S., Hollweg, J.~V., \& Esser, R. 1999, J.\ Geophys.\ Res., {\bf
  104}, 2521

\bibitem[{{Mangeney} {et~al.}(1991){Mangeney}, {Grappin}, \&
  {Velli}}]{Mangeney_al_1991}
{Mangeney}, A., {Grappin}, R., \& {Velli}, M. 1991, Advances in Solar System
  Magnetohydrodynamics ({Priest}, E. R. and {Hood}, A. W.), 327

\bibitem[{{Matthaeus} {et~al.}(1983){Matthaeus}, {Montgomery}, \&
  {Goldstein}}]{Matthaeus_al_1983}
{Matthaeus}, W.~H., {Montgomery}, D.~C., \& {Goldstein}, M.~L. 1983, Physical
  Review Letters, {\bf 51}, 1484

\bibitem[{{Matthaeus} {et~al.}(1998){Matthaeus}, {Smith}, \&
  {Oughton}}]{Matthaeus_al_1998}
{Matthaeus}, W.~H., {Smith}, C.~W., \& {Oughton}, S. 1998, J.\ Geophys.\ Res.,
  {\bf 103}, 6495

\bibitem[{{Matthaeus} {et~al.}(1999){Matthaeus}, {Zank}, {Oughton}, {Mullan},
  \& {Dmitruk}}]{Matthaeus_al_1999}
{Matthaeus}, W.~H., {Zank}, G.~P., {Oughton}, S., {Mullan}, D.~J., \&
  {Dmitruk}, P. 1999, ApJL, {\bf 523}, L93

\bibitem[{{Matthaeus} {et~al.}(1994){Matthaeus}, {Zhou}, {Zank}, \&
  {Oughton}}]{Matthaeus_al_1994}
{Matthaeus}, W.~H., {Zhou}, Y., {Zank}, G.~P., \& {Oughton}, S. 1994, J.\
  Geophys.\ Res., {\bf 99}, 23421

\bibitem[{{Munro} \& {Jackson}(1977)}]{Munro_Jackson_1977}
{Munro}, R.~H. \& {Jackson}, B.~V. 1977, Astrophys.\ J., {\bf 213}, 874

\bibitem[{{Oughton} {et~al.}(2004){Oughton}, {Dmitruk}, \&
  {Matthaeus}}]{Oughton_al_2004}
{Oughton}, S., {Dmitruk}, P., \& {Matthaeus}, W.~H. 2004, Physics of Plasmas,
  {\bf 11}, 2214

\bibitem[{{Oughton} {et~al.}(2001){Oughton}, {Matthaeus}, {Dmitruk}, {Milano},
  {Zank}, \& {Mullan}}]{Oughton_al_2001}
{Oughton}, S., {Matthaeus}, W.~H., {Dmitruk}, P., {et~al.} 2001, Astrophys.\
  J., {\bf 551}, 565

\bibitem[{{Pruneti} \& {Velli}(1997)}]{Pruneti_Velli_1997}
{Pruneti}, F. \& {Velli}, M. 1997, in ESA SP-404: Fifth SOHO Workshop: The
  Corona and Solar Wind Near Minimum Activity, 623--+

\bibitem[{{Similon} \& {Zargham}(1992)}]{Similon_Zargham_1992}
{Similon}, P.~L. \& {Zargham}, S. 1992, Astrophys.\ J., {\bf 388}, 644

\bibitem[{Smith {et~al.}(2001)Smith, Matthaeus, Zank, Ness, Oughton, \&
  Richardson}]{Smith_al_2001}
Smith, C.~W., Matthaeus, W.~H., Zank, G.~P., {et~al.} 2001, J.\ Geophys.\ Res.,
  {\bf 106}, 8253

\bibitem[{{Suzuki} \& {Inutsuka}(2005)}]{Suzuki_Inutsuka_2005a}
{Suzuki}, T.~K. \& {Inutsuka}, S.-i. 2005, ApJL, {\bf 632}, L49

\bibitem[{{Tu}(1988)}]{Tu_1988}
{Tu}, C.-Y. 1988, J.\ Geophys.\ Res., {\bf 93}, 7

\bibitem[{{Tu} {et~al.}(1984){Tu}, {Pu}, \& {Wei}}]{Tu_al_1984}
{Tu}, C.-Y., {Pu}, Z.-Y., \& {Wei}, F.-S. 1984, J.\ Geophys.\ Res., {\bf 89},
  9695

\bibitem[{{Velli}(1993)}]{Velli_1993}
{Velli}, M. 1993, Astron.\ Astrophys., {\bf 270}, 304

\bibitem[{{Velli} {et~al.}(1989){Velli}, {Grappin}, \&
  {Mangeney}}]{Velli_al_1989}
{Velli}, M., {Grappin}, R., \& {Mangeney}, A. 1989, Physical Review Letters,
  {\bf 63}, 1807

\bibitem[{{Velli} {et~al.}(1991){Velli}, {Grappin}, \&
  {Mangeney}}]{Velli_al_1991}
{Velli}, M., {Grappin}, R., \& {Mangeney}, A. 1991, Geophys. Astrophys. Fluid
  Dynamics, {\bf 62}, 101

\bibitem[{{Verdini} {et~al.}(2005){Verdini}, {Velli}, \&
  {Oughton}}]{Verdini_al_2005}
{Verdini}, A., {Velli}, M., \& {Oughton}, S. 2005, Astron.\ Astrophys., {\bf
  444}, 233

\bibitem[{{Wilhelm} {et~al.}(1995){Wilhelm}, {Curdt}, {Marsch}, {Schuhle},
  {Lemaire}, {Gabriel}, {Vial}, {Grewing}, {Huber}, {Jordan}, {Poland},
  {Thomas}, {Kuhne}, {Timothy}, {Hassler}, \& {Siegmund}}]{Wilhelm_al_1995}
{Wilhelm}, K., {Curdt}, W., {Marsch}, E., {et~al.} 1995, Solar Phys., {\bf
  162}, 189

\bibitem[{{Zank} {et~al.}(1996){Zank}, {Matthaeus}, \& {Smith}}]{Zank_al_1996}
{Zank}, G.~P., {Matthaeus}, W.~H., \& {Smith}, C.~W. 1996, J.\ Geophys.\ Res.,
  {\bf 101}, 17093

\bibitem[{{Zhou} \& {Matthaeus}(1989)}]{Zhou_Matthaeus_1989}
{Zhou}, Y. \& {Matthaeus}, W.~H. 1989, Geophys.\ Rev.\ Lett., {\bf 16}, 755

\bibitem[{{Zhou} \& {Matthaeus}(1990)}]{Zhou_Matthaeus_1990}
{Zhou}, Y. \& {Matthaeus}, W.~H. 1990, J.\ Geophys.\ Res., {\bf 95}, 10291

\end{thebibliography}
\end{document}